\documentclass[10pt]{article}
\usepackage[OE]{express}
\usepackage{siunitx}

\begin{document}
\title{Inner structure of ZnO microspheres fabricated via laser ablation in superfluid helium}

\author{Yosuke Minowa,\authormark{1,*} Yuya Oguni,\authormark{1} and Masaaki Ashida\authormark{1}}

\address{\authormark{1}Graduate School of Engineering Science, Osaka University, Toyonaka, Osaka 560-8531, Japan}

\email{\authormark{*}minowa@mp.es.osaka-u.ac.jp} 



\begin{abstract}
ZnO microspheres fabricated via laser ablation in superfluid helium were found to have bubble-like voids. Even a microsphere demonstrating clear whispering gallery mode resonances in the luminescence had voids. Our analysis confirmed that the voids are located away from the surface and have negligible or little effect on the whispering gallery mode resonances since the electromagnetic energy localizes near the surface of these microspheres. The existence of the voids indicates that helium gas or any evaporated target material was present within the molten microparticles during the microsphere formation. 
\end{abstract}




\section{Introduction}

Dielectric microspheres have the ability to confine light to a small volume with a high quality factor through internal total reflection forming the electromagnetic eigen modes known as whispering gallery modes (WGMs)\cite{buck_optimal_2003}. The strong coupling between light and matter offers a variety of applications such as microlasers\cite{spillane_ultralow-threshold_2002}, enhanced nonlinear optical devices\cite{treussart_evidence_1998}, and cavity quantum electrodynamics' platform\cite{buck_optimal_2003}. The light-matter coupling can be naturally resulted by fabricating microspheres from materials having large oscillator strength. Excitons in direct band gap semiconductors have large oscillator strengths\cite{t_hooft_giant_1987} and high luminescence quantum yields, which make them suitable to be coupled with the microcavity optical modes.

The fabrication of semiconductor microspheres remains challenging\cite{nagai_spherical_1997,ngo_size_2016} owing to their crystalline structure. This can be compared with the ease of fabrication of the amorphous microspheres of silica or polymer\cite{stober_controlled_1968,gorodetsky_ultimate_1996}. As the symmetry of the crystal structure determines the thermodynamically favored shape of the materials, the crystal with slow growth rate results in a non-spherical shape. Laser ablation is one of the widely used methods to fabricate micro- and nanoparticles from various materials including semiconductors\cite{semaltianos_nanoparticles_2010}, which is the reverse of the slow crystal growth. In particular, nanosecond-pulsed laser ablation in superfluid helium can produce semiconductor microspheres with high symmetry and smooth surface\cite{okamoto_optical_2014,okamoto_fabrication_2014}. The fabricated microspheres show a crystalline structure even at the surface and at remarkably low threshold WGM lasing with continuous-wave laser excitation\cite{okamoto_white-light_2012}. The laser ablation in superfluid helium can also produce metallic microspheres\cite{gordon_stability_2014} and semiconductor nanoparticles\cite{inaba_optical_2006}. An in-depth investigation of the microscopic fabrication mechanism is necessary to open the possibilities of the method targeting the microsphere fabrication  with different materials under different conditions\cite{nakamura_synthesis_2013}. The observation of the inner structure of the fabricated microspheres would provide essential insight into the fabrication mechanism. Although semiconductor spheres with a size $\lesssim$ \SI{300}{\nano \meter} fabricated via the same method had very few dislocations or defects and were proved to be single crystals by transmission electron microscopy (TEM)\cite{okamoto_white-light_2012}, the investigation of the detailed inner structure of the microspheres with a size of $\gtrsim$ \SI{1}{\micro \meter} is difficult owing to limited electron beam penetration depth.

Here, we have demonstrated that the semiconductor ZnO microspheres fabricated via laser ablation in superfluid helium contain bubble-like voids with a size from a few tens of nanometers to sub-micrometers. Through experiments, we also proved that the microspheres with voids can maintain WGM resonances. Based on the calculation of light intensity distribution within the microsphere, it can be assumed that the a void at the center of the microsphere can have little effect on WGMs. Furthermore, all the observed microspheres contained voids, although their sizes and the locations were different. The presence of a large number of voids indicates that helium gas or the gas phase of any ablated material may play some role during the formation of the microspheres after laser ablation. Moreover, our findings suggest that the size and location of the voids are some hidden parameters that explain the differences among the fabricated microspheres of the quality factor of the WGMs\cite{okamoto_fabrication_2014}.

\section{Experiment}
A sintered semiconductor ZnO target with a diameter of 10 mm and a thickness of 3 mm was placed on a sample holder in a cryostat filled with the superfluid helium. ZnO is a direct band gap semiconductor material with a large band gap energy of \SI{3.4}{\eV}. Excitons can be formed by the absorption of light or a beam of electrons even at room temperature because of large exciton binding energy, \SI{60}{\milli \eV}. The exciton formation ensures a high luminescence quantum yield. The target surface was irradiated with the second harmonic of Nd:YAG laser with a pulse duration of 10 ns, a pulse energy of 1 mJ, and a repletion rate of 10 Hz. The spot size at the target surface was around \SI{50}{\micro \meter} with the focusing lens $f = $ \SI{200}{\milli \meter}. At the bottom of the sample holder, we placed a substrate made of Si to accumulate the fabricated particles for further analysis. We examined the microscopic morphology of the fabricated microspheres by using a scanning electron microscope (SEM). Furthermore, cathodoluminescence spectroscopy was performed to ensure the quality of the microspheres as an optical microcavity through the observation of the WGM resonances in the luminescence. Then, we examined the cross section of the fabricated microspheres after the focused ion beam (FIB) processing. Finely focused ion beams milled and cut the sample with nanometer scale precision. We observed the shape of FIB-processed microspheres in the perpendicular direction and at 38$^\circ$ oblique to the cross-sectioned surface normal.

\section{Result and discussion}
Figures 1(a) and (d) show the SEM images of a typical ZnO microsphere fabricated with laser ablation in superfluid helium. The two images were taken for the same microsphere from two different angles. The shape was observed to be highly spherical and the surface was smooth, although there were several dark lines indicating possible microscopic structural boundaries. We also observed the cathodoluminescence from the microsphere as shown in Fig. 2. The spectrum shows ultraviolet emission that originates from the exciton in ZnO and broad visible emission which is thought to be a defect-related luminescence\cite{ozgur_comprehensive_2005}. We can observe many peaks corresponding to the WGM resonances validating the fact that the microsphere behaves as a good optical microcavity. 

\begin{figure}[htbp]
\centering\includegraphics[width=7cm]{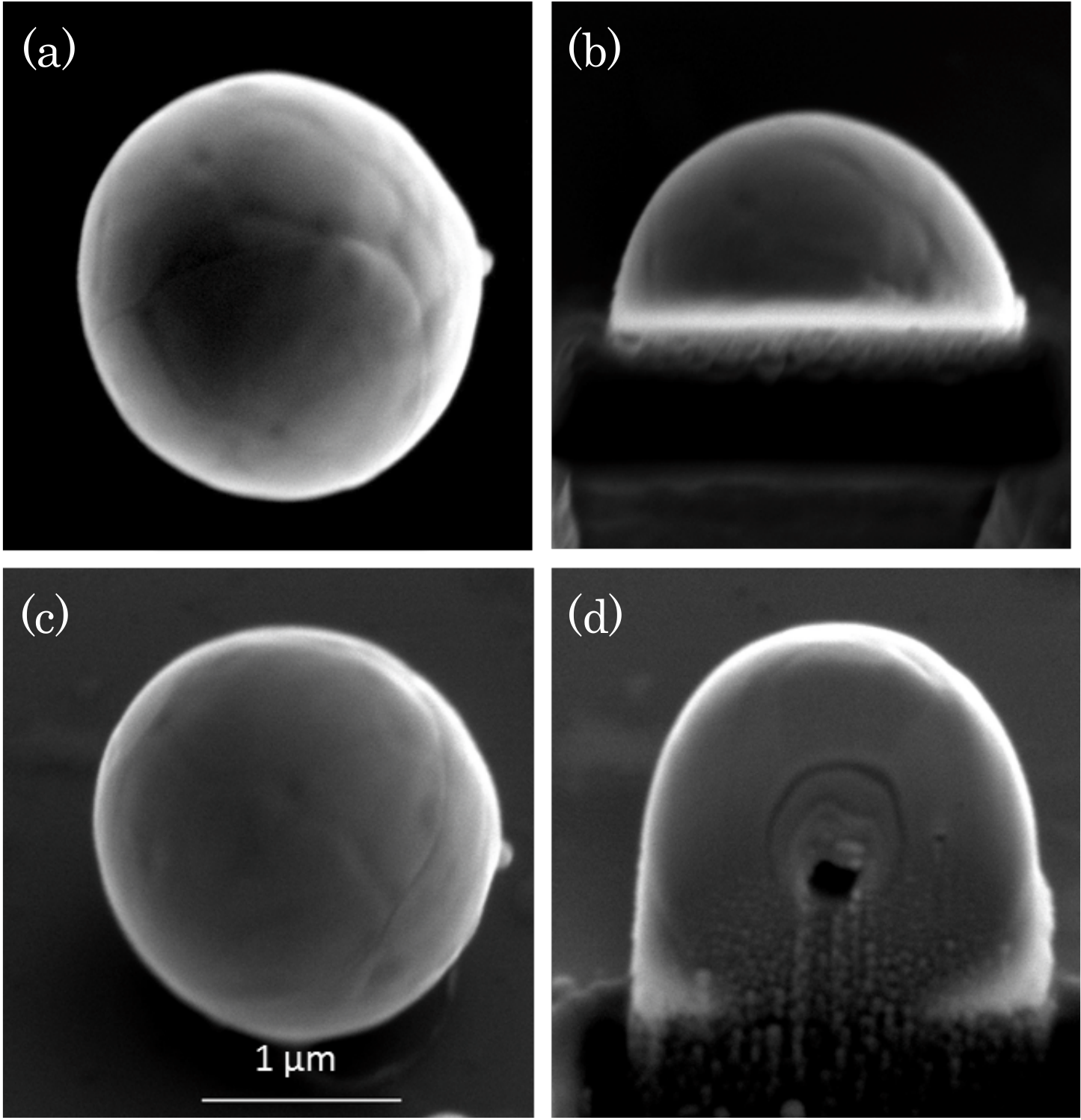}
\caption{SEM images of a ZnO microsphere fabricated via laser ablation in superfluid helium before (a and c) and after (b and d) the FIB cross-sectioning. The microphere was observed from the  perpendicular direction (a and b) or  38$^\circ$ oblique (c and d) to the cross-sectioned surface normal.}
\end{figure}
We calculated WGM resonance wavelengths based on the Mie scattering theory\cite{bohren_absorption_1983} with the frequency-dependent refractive index\cite{nobis_low-order_2005}
\begin{equation}
n_\mathrm{material}(h\nu)=1.916+1.145\times10^{-2}\left(h\nu\right)^2+1.6507\times10^{-3}\left(h\nu\right)^4,
\end{equation}
where $h\nu$ is the photon energy in eV and a diameter of \SI{1.033}{\micro\meter}, which are found to be consistent with the size of the microsphere estimated from the SEM images. The dotted lines shown in Fig. 2 correspond to the transverse electric (TE) WGM resonance wavelengths with a radial mode number $n=1$, which have intrinsically higher quality factor than the other modes\cite{datsyuk_characteristics_1992}. The polar mode numbers are denoted at the top of the graph. The calculated WGM resonance wavelengths describe well the experimentally derived peaks. 

\begin{figure}[htbp]
\centering\includegraphics[width=8cm]{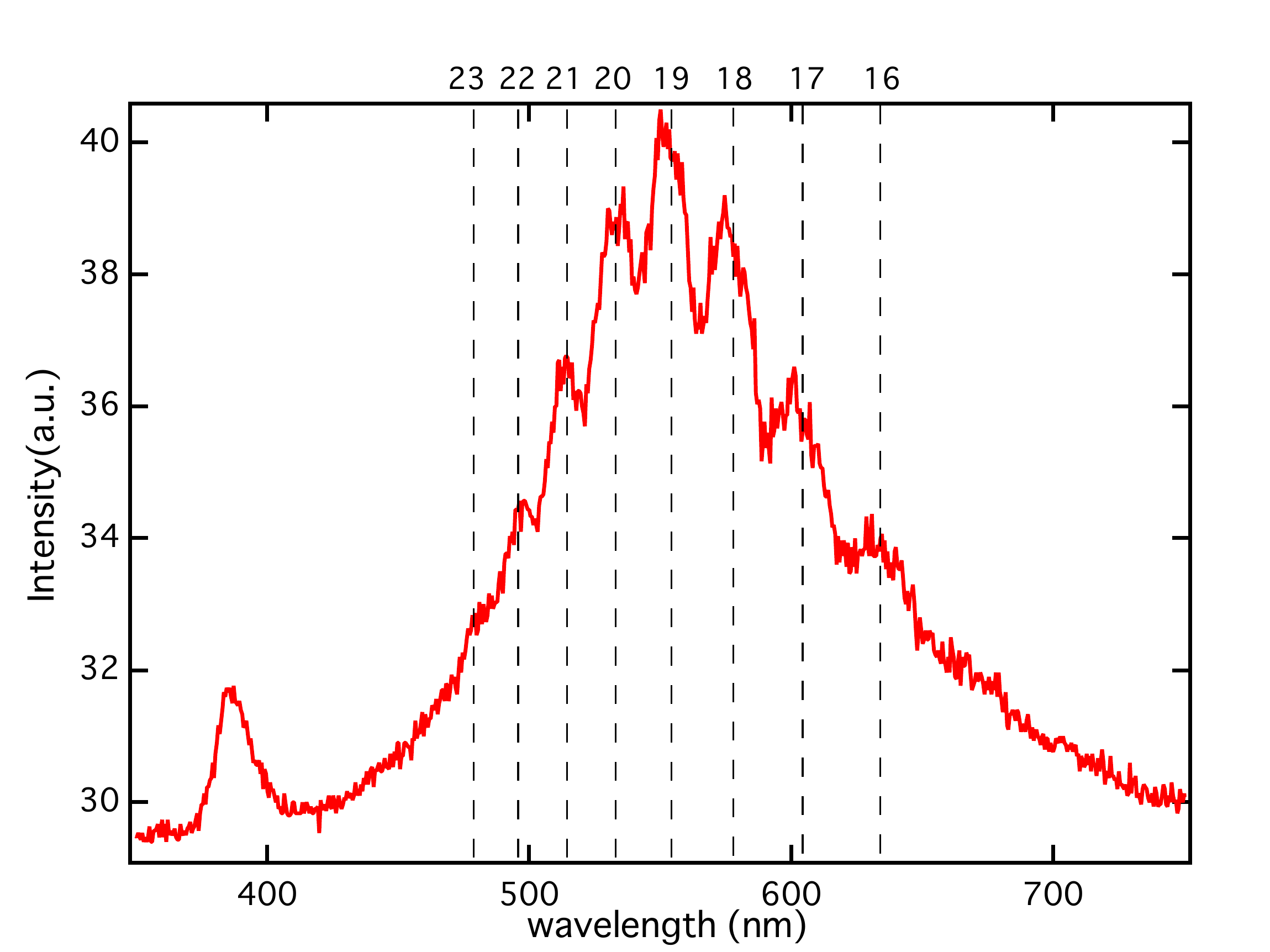}
\caption{CL spectrum from the ZnO microsphere shown  in Fig. 1, before the FIB cross-sectioning. The dotted lines correspond to the WGM resonance wavelengths estimated from Mie scattering theory. The numbers above the graph indicate the polar mode number $l$.}
\end{figure}

For further analysis, we observed a cross section of the microsphere by using both FIB and SEM (Fig. 1(d)), which lies approximately at the middle of the microsphere as shown in Fig. 1(b). We clearly observe a large void at the core of the microsphere and a small one right next to the large void, although redeposition of sputtered material obscured slightly the cross section. At first glance, the existence of the voids seems to contradict the WGM mode resonances in the luminescence as the voids in the microsphere are strong light scatterers. To clarify the contradictory findings, we calculated the electric field intensity distribution\cite{oraevsky_whispering-gallery_2002} within the microsphere for a typical TE WGM mode with a radial mode number $n=1$, a polar number $l=19$, an azimuthal mode number $m=l$, and a wavelength of \SI{554.31}{\nano \meter} corresponding to the highest peak shown in Fig. 2. The energy of the electromagnetic wave localizes only at the surface. We can thus neglect the effect of the voids on the WGM resonances if the voids locate far away from the microsphere's surface. The exact layer width of the WGM energy localization in the radial direction is given by\cite{datsyuk_optics_2001,datsyuk_characteristics_1992}
\begin{equation}
\dfrac{|t_n|\left[ \dfrac{1}{2}\left(l+\dfrac{1}{2}\right) \right]^{1/3}\lambda}{2\pi n_\mathrm{material}},
\end{equation}
where $t_n$ is the $n$th zero of the Airy function and $\lambda$ is the vacuum wavelength. With our experimental parameters, the energy localization width for $n=1$ mode is calculated to be \SI{200}{\nano \meter} $\sim$ \SI{240}{\nano \meter} depending on the polar mode number. The small width ensures that the WGM energy is localized in the region without the voids shown in Fig. 1(d). The results also confirm that the location of these voids is an important parameter affecting the scattering loss of the WGMs.

\begin{figure}[htbp]
\centering\includegraphics[width=7cm]{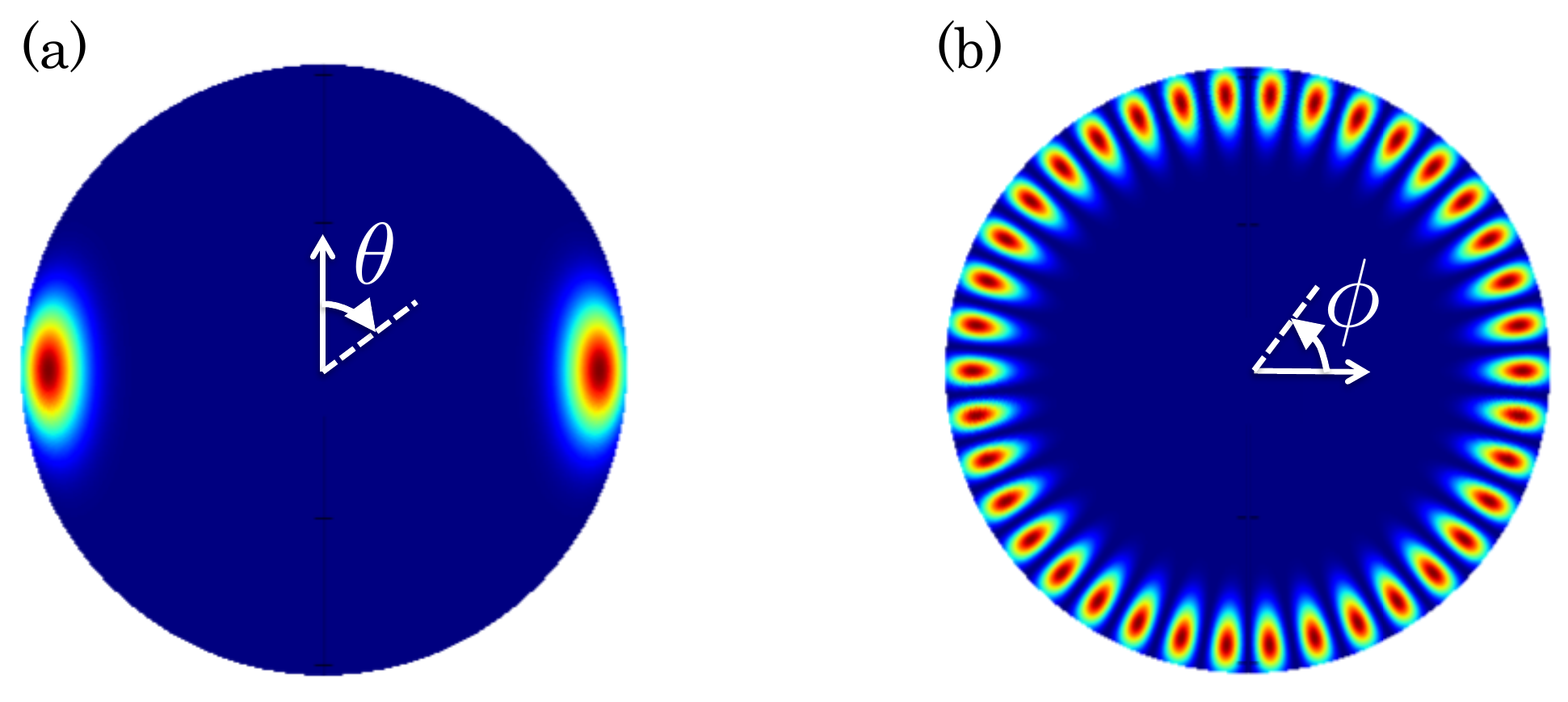}
\caption{Calculated WGM electric field intensity distribution within the ZnO microsphere with a radius of \SI{1.04}{\micro\meter}, a radial mode number $n=1$, a polar mode number $l=19$ and an azimuthal mode number $m=l$. Polar (a) and azimuthal (b) distributions.}
\end{figure}

All the observed microspheres fabricated via the laser ablation in superfluid helium contained voids, although the sizes and the relative locations of the voids were not the same. Figure 4 shows the serial cross-sectioned images of the microsphere with the largest void observed. Figure 4(a)-(f) are the images observed from the direction perpendicular to the cross-sectioned surface normal and Fig. 4(g)-(l) show the images observed from the direction 38$^\circ$ oblique to the cross-sectioned surface normal. The cross-section observation confirmed an almost uniformly thin spherical shell structure of the microsphere. Furthermore, the structure consisted of many polygons, each of which may correspond to a microcrystal. The existence of large voids suggests that the helium gas generated by instantaneous heating of the target material or the gas phase of the ablated material plays an important role during the formation of the microsphere after the laser ablation. The presence of the gas within the molten ZnO during the solidification may explain the void formation. The formation of voids with smooth spherical surface indicates that the solidification process is considerably fast as suggested in the simulation for the formation of a certain type of chondrules (spherical grains found in chondrites) \cite{miura_new_2011}. 

\begin{figure}[htbp]
\centering\includegraphics[width=12cm]{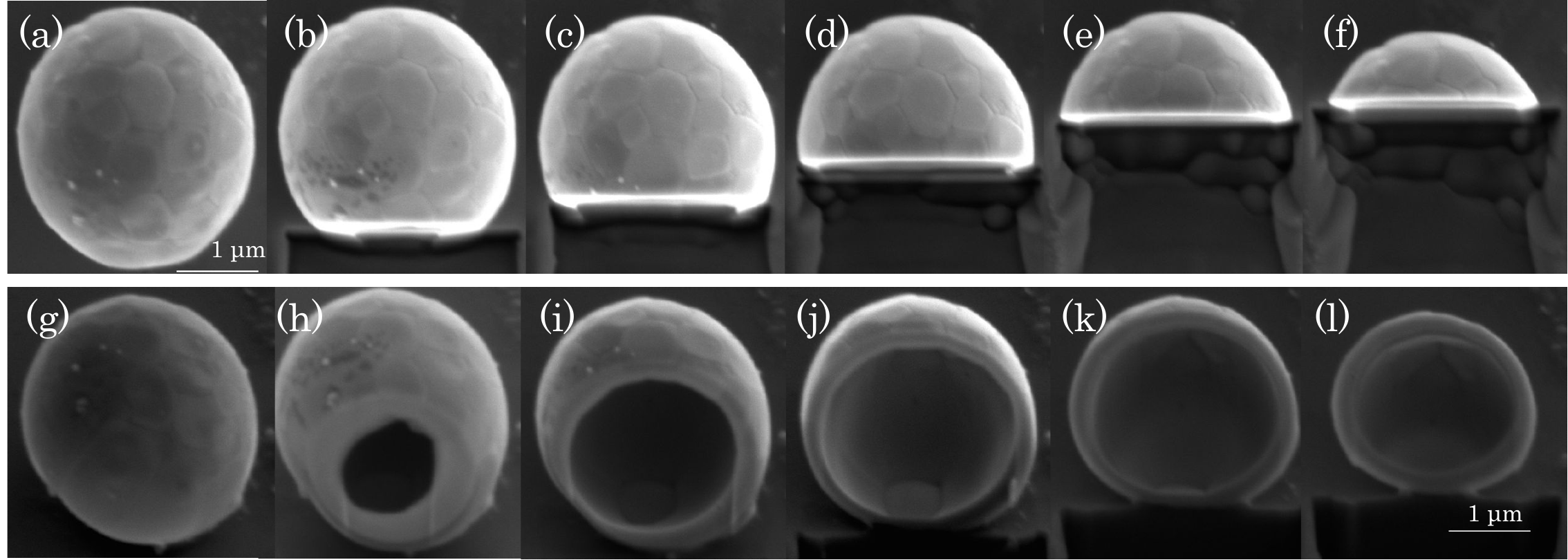}
\caption{Serial cross-sectioned images of a ZnO microsphere fabricated via the laser ablation in superfluid helium.  The microsphere was observed from the perpendicular direction  (a-f) or  38$^\circ$ oblique (g-l) to the cross-sectioned surface normal.}
\end{figure}

\section{Conclusion}
We found that the ZnO microspheres fabricated via laser ablation in superfluid helium contain some voids. The existence of the voids may affect the WGM resonances although the WGMs can be unaltered if the voids locate far away from the surface of the microsphere where the electromagnetic energy localizes. The location of the voids is one of the important parameters determining the loss of the WGMs and the optical properties of the fabricated microspheres. The voids could be the result of the inclusion of the helium gas or the gas phase of the ablated material within the molten ZnO particles. 
Large-scale fabrication of the microspheres with high quality factors from various materials requires suppressing the void formation, which may be controlled by changing the ablation conditions such as the pulse energy and the photon energy. The formation of the microspheres without voids would be possible since spheres with a size of $\lesssim$ \SI{300}{\nano \meter} fabricated via the same method already had very a few dislocations or defects\cite{okamoto_white-light_2012}. If we can control the formation of such voids, We can apply the laser ablation method also to produce the homogeneous hollow semiconductor microspheres, which have potential in gas sensing and photocatalytic devices owing to their large surface-to-volume ratio\cite{ihara_template-free_2014} and are interesting targets for the optical vortex trapping\cite{gahagan_optical_1996}.

\section*{Funding}
JSPS KAKENHI Grant Number JP15K13501, JP16H06505, JP16H03884.; The Murata Science Foundation.; the Izumi Science and Technology Foundation.

\section*{Acknowledgments}
The authors wish to thank Satoshi Ichikawa for the technical assistance with FIB and SEM observation. Y. M. is grateful to Hiromasa Niinomi for fruitful discussions.

\end{document}